\documentclass[runningheads]{llncs}
\usepackage{graphicx}
\usepackage{cite,url,booktabs}
\begin{document}
\title{A novel dataset for the identification of computer generated melodies in the CSMT challenge\thanks{Supported by Zhongwen Law Firm. Shengchen Li and Yinji Jing are considered as the co-first authors.}}
\titlerunning{A dataset for CSMT data challenge}
%
\author{Shengchen Li\inst{1}\orcidID{0000-0002-2488-298X} \and
Yinji Jing\inst{1} \and
Gy\"orgy Fazekas\inst{2}\orcidID{0000-0003-2580-0007}}
\authorrunning{S. Li, Y. Jing and G. Fazekas}
%
\institute{Beijing University of Posts and Telecommunications\\ \email{\{shengchen.li,jyj\}@bupt.edu.cn} \and
Queen Mary University of London, London, UK\\
\email{g.fezekas@qmul.ac.uk}}
\maketitle              
\begin{abstract}
This paper introduces a novel dataset for the identification of computer generated melodies as used in the data challenge organised by the Conference on Sound and Music Technology (CSMT).
The CSMT data challenge requires participants to identify whether a given piece of melody is generated by computer or is composed by human. The dataset consists of two parts: a development dataset and an evaluation dataset. The development dataset contains only computer generated melodies whereas the evaluation dataset contain both computer generated melodies and human composed melodies. The aim of the dataset is to facilitate the develpment and assessment of methods to identified computer generated melodies and facilitate the creation of generative music systems.

\end{abstract}
\section{Introduction}
Automatic music generation is becoming more and more popular with the development of deep learning techniques. At the same time, new challenges have emerged in juridical practices regarding copyright protection: the source of music leads to different juridical models. Although the discussion of legal issues is beyond the scope of this paper and it isn't among the aims of the proposed data challenge, a new task is considered helpful in future juridical practices. Identifying whether a piece of melody is computer generated or human composed could help in recognising cases of music use where legal intervention of further scrutiny is necessary. As a result, the Conference on Sound and Music Technology (CSMT) proposes a data challenge that requires participants to identify human composed melodies among computer generated ones.

Existing automatic music generation methods have a certain drawbacks, such as the lack of clear long term structure in the music or the existence of unusual harmonisation, which make the melody identification less challenging. For example, SSM is used to identify the repetitions in music \cite{cheng2018music} that are commonly associated to music structure by composers but are seldom generated by music generation algorithms \cite{herremans2017functional}. Moreover in juridical practices, copyright infringement can be detected using the similarity of the variation of pitch in melodies, regardless of music structure and accompaniment. In this paper and the proposed challenge, the term ``melody'' refers to a sequence of pitches with dedicated duration but excludes the concept of music structure and accompaniment. 

The proposed data challenge follows a possible scenario of melody source identification in juridical practice. There are two datasets used in the challenge. The development dataset consists of computer-generated melodies that are produced by a set of exemplar music generation systems. The evaluation dataset contains both computer-generated and human-composed melodies. Participants are required to submit a system that identifies human-composed melodies among the computer-generated melodies. 

The authors and organisers of the data challenge reviewed existing computer music generation systems as outlined in this paper. Three exemplar methodologies were proposed including Generative Adversarial Network (GAN), Variational Auto Encoder (VAE) and transformer systems because these architectures are commonly used and represent the current state-of-the-art in music generation as of early 2020. All systems were used to produce computer-generated melodies in both development and evaluation datasets. The systems used to generate melodies for development and evaluation datasets are different as the results of a different initial values and different batch formation in the training process. 
For human-composed melodies in the evaluation dataset, the majority (95\%) of human-composed melodies within the evaluation dataset overlaps with human-composed melodies used as training data for the automatic generation system. The remaining human-composed melodies are composed by university students whose major is music composition. Such melodies have not been published to the public.

The proposed data challenge can be approached in two different ways. If human-composed melodies are collected by the participants, data may be labelled as ``human'' vs. ``computer'', hence the proposed task can be considered a binary classification problem. The human-composed melodies can also be considered outliers among computer-generated melodies. In this case, the proposed task can also be viewed as an unsupervised outlier detection problem.

The rest of the paper is organised as follows. A brief overview of automatic music generation is presented in Section 2 in order to justify the choice of melody generation systems. In Section 3, the dataset creation process is explained in detail together with the data representation proposed for the challenge. This is followed by a brief conclusion in Section 4.

\section{Melody Generation Systems}

This section provides an overview of automatic music generation systems. The majority of music generation systems can be divided to three types \cite{liu2016computational}: rule-based systems, methods that utilise mathematical models and machine learning systems. The machine learning systems, especially deep learning systems, are considered as the state-of-the-art automatic music generation systems  \cite{briot2020deep}. As a result, the data challenge proposes to use deep learning systems to generate melodies that are labelled as computer-generated melodies.

The most important factor for automatic music generation that affects system performance is the modelling of temporal dependencies. Rule-based systems usually propose a set of rules to generate a sequences such as chords \cite{steedman1984generative}. Systems that use mathematical models aim to describe time dependency in music mathematically. The generation process may then be considered a sampling process from a mathematical model. For modelling temporal dependencies, Markov models are considered the first choice since the very early stages of music generation \cite{pinkerton1956information}. One of more the recent works using this principle is the ALYSIA system \cite{ackerman2017algorithmic} that creates both lyrics and melodies. 

As music usually has a long-time dependency, it is almost impossible for rule-based and mathematical modelling systems to learn long-time dependencies accurately. Machine learning systems especially deep learning systems are better suited for the purpose of music generation as the long-term dependency can be modelled as a joint probability distribution akin to a language model \cite{eck2002first}. 

One exemplar system is a Recurrent Neural Network (RNN). Makris  \cite{makris2017combining} uses RNN to generate rhythm in drum patterns. The Microsoft team \cite{zhu2018xiaoice} uses RNN to encode the pitch, rhythm and chord of music. With the development of transformer systems that are better at modelling longer-time dependencies, Vaswani et al. \cite{vaswani2017attention} proposed transformer structure to catch longer temporal-dependency. This was adopted by Huang et al. \cite{huang2018music} for music generation. In the proposed data challenge, the MusicTransformer \cite{huang2018music} system is used as one of the candidate system to generate computer-generated melodies, where the authors claimed that the MusicTransformer models long-term dependencies in music \cite{huang2018music}. 

Besides using a language model to model long-time dependency in music, music generation can also be performed by a generative model such as a Variational Auto-Encoder (VAE) or a Generative Adversarial Network (GAN).

VAE is a variant of the autoencoder, which is a generative deep learning model. Brunner \cite{brunner2018symbolic} proposed a VAE-based automatic composition model MIDI-VAE, which processes polyphonic music with multiple instrument tracks and models the duration and speed of the notes in the generated music. Wang \cite{wang2019modeling} proposed a new variant of Variational Autoencoder (VAE), which uses a modular approach to designing the model structure to generate music. Luo \cite{luo2020mg} used a variational autoencoder to generate different styles of Chinese folk music. MusicVAE \cite{pmlr-v80-roberts18a} improves the structure of VAE according to the characteristics of music with hierarchical structure, which aims to solve the lack of coherence in generated music using vanilla VAE. The musicVAE system is better at generating music with extended duration hence the proposed data challenge selects musicVAE as the representative of VAE-based music generation systems in the development and evalution datasets. 

Generative adversarial network (GAN) \cite{goodfellow2014generative} is a generative model that contains a generator and a discriminator. In a GAN, the generator produces pseudo-samples and the discriminator judges whether a sample was  produced by the generator. GAN is commonly used for music generation, for example, by Liu and Yang \cite{liu2018lead} and Dong et al. \cite{dong2018musegan}.  MidiNet \cite{yang2017midinet} is one of few GAN systems that use piano roll as the representation of music and can generate melodies without the generation of music accompaniment. As a result, the proposed data challenge selects MidiNet as the choice of GAN based systems for music generation.

To summarise, deep learning based computer music generation systems outperform conventional rule-based and mathematical modelling systems. Among deep learning systems, there are three types of systems that are considered state-of-the-art: transformer systems, VAE-based systems and GANs. The proposed data challenge selects an exemplar system to represent each of these types: MusicTransformer, MusicVAE and MidiNet (GAN). The computer-generated melodies in the development and evaluation datasets are a combination of melodies generated by all three selected systems.

\section{Dataset}
\subsection{Training Data}

To investigate whether different music style affects the identification of computer-generated melodies, two datasets are used for training the selected models: Bach Chorales in Music21\footnote{\url{https://web.mit.edu/music21/}} and pop music from hooktheory\footnote{\url{https://www.hooktheory.com/}}. These two training datasets are used for training two separate models for melody generation in this data challenge. 

The raw melodies in the datasets are subject to a pre-processing stage. The Bach Chorales dataset contains several voices. Each voice is treated as a separate melody. With regards to pop music in hooktheory, only the melody part is used for training. All melodies are truncated to 32 beats to disregard music structure.

\begin{figure}[htb]
	\centering
	\includegraphics[width=\textwidth]{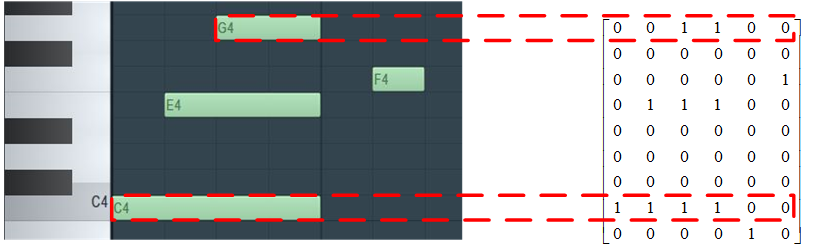}
	\caption{Binarised Piano Roll Representation}
	\label{fig:piano_roll}
\end{figure}

As used by all selected systems \cite{yang2017midinet, pmlr-v80-roberts18a, huang2018music}, all pre-processed melodies for training are converted into a form of binarised piano roll as demonstrated in Fig.~\ref{fig:piano_roll}. The binarised piano roll represents melodies using a matrix, where each column represents a quarter beat and each row represents a note (such as A4). As each melody has a length of 32 beats and each column represents a quarter beat, the binarised piano roll has 128 ($32\times 4 = 128$) columns. Moreover, as MIDI files have a pitch number defined between \texttt{0} to \texttt{127}, there are 128 rows in the binarised piano roll. As a result, the music representation in this paper has a shape of $128\times 128$.

\subsection{Computer-generated Melodies}
As discussed in our brief overview of music generation methods, the selected systems for melody generation are MusicTransformer \cite{huang2018music}, MusicVAE \cite{pmlr-v80-roberts18a} and MidiNet \cite{yang2017midinet}. In this section, the working principles of these systems are outlined briefly. For more details, the reader is kindly asked to refer to the original papers.

Two datasets (\textit{Music21} and \textit{hooktheory}) are used to train all selected systems twice under the exact configuration hence two models are obtained for each style: Bach and Pop. For each style, one of the resulting models are used to generate melodies in the development dataset, the remaining model is used to generate melodies in the evaluation dataset. 

\subsubsection{MusicTransformer}

MusicTransformer \cite{huang2018music} uses a Neural Network Language Model (NNLM) to generate music where the pitch and duration of notes at a time can be considered a word and the motives or phrases can be considered a sentence. This work is among the first using a Transformer to generate music.

\sloppy
Given a sentence $S$ which contains $N$ words $w_i$, that is, $S=<w_1, w_2,...,w_n> \in V_n$, $V_n$  is the size of the overall vocabulary. The language model aims to find the probability distribution of the sentence, which can be formalised using the equation (\ref{eq:1}). Given the forward sequence of a word, the probability of the entire word sequence can be decomposed into the product of the conditional probability of the next word with respect to its forward word. The results of the system show that longer temporal dependencies are well modelled, since repeated or similar phrases can be found in the music generated by the proposed system. 

\begin{equation}
	P(S)=P(w_1, w_2,...,w_N)=P(w_1)P(w_2|w_1)\cdot \cdot \cdot P(w_N|w_1, w_2,...,w_{N-1})
	\label{eq:1}
\end{equation}

\fussy
The initialisation process of the system depends on the joint probability distribution of the initial sequences hence usually a randomly selected melody with dedicated lengths is used for initialisation. In this data challenge, the effects of the initialisation process for the MusicTransformer are also investigated by examining whether melodies generated by different initialisation seeds can be identified.

\subsubsection{MusicVAE}

MusicVAE \cite{pmlr-v80-roberts18a} improves the structure of VAE according to the characteristics of music with hierarchical structure, which aims to solve the problem of lack coherence in the generated music when a vanilla VAE is used. The music is first represented using an encoder, constructed with a recurrent neural network to obtain a low-dimensional hidden vector. The resulting vector is then decoded with a multi-level decoder, which reconstructs the vector into a 16-bar unit first, then the decoding process continues with lower-level decoders to generate finer units of melodies.

\subsubsection{MidiNet}
MidiNet \cite{yang2017midinet} converts music binarised into a piano roll, which is akin to a two-dimensional image. The generator and discriminator of the GAN system then use convolutional neural networks to encode and decode the resulting binarised piano rolls. Besides binarised piano rolls generated by decoders, music composed by humans is also sent to the discriminator for training. At the same time, to maintain coherent connection between the music segments, MidiNet adds information of the front music segment to each layer in the generator. This system is one of the earliest works targeting automatic composition using the method of generating images. It demonstrates the feasibility of using CNN to generate piano roll.

\subsection{Human-composed Melodies}
Human composed melodies in this challenge have two sources. Published melodies that are used train the selected music generation systems and unpublished melodies that are required to be composed for this data challenge by university students majoring in music composition. 

Published melodies are randomly selected from the dataset that trains the selected music generation systems. The selected melodies are then truncated to 32-beat long segments. 

The unpublished melodies in the evaluation dataset is used to test the ability of recognising unknown human-composed music. The data challenge invited professionally trained composers from the China Conservatory of Music to compose a number of melodies. The students were asked to compose melodies in two styles: the Baroque style as composed by J. S. Bach and the common pop style. The structure of the composed melodies is removed with melody truncated to 32-beat long as well.

\subsection{Data Representation}
 
The paper uses \texttt{pretty\_midi}\footnote{\url{http://craffel.github.io/pretty-midi/}} to convert the generated piano roll into a MIDI file which requires the MIDI number and the duration of each note. The MIDI number can be directly indexed by the note. The duration of each beat requires a simple calculation. As each column in the binarised piano roll represents a quarter beat, given a tempo value, such as 120 beats per minute (bpm), the duration of each column in the binarised piano roll can be easily calculated.

The instrument selected in the MIDI file is ``\texttt{Bright\_Piano}'' with the velocity setting to 127 in MIDI files. The tempi of the MIDI files are randomly selected in the range of 68 bpm, 78 bpm, 88 bpm, 98 bpm, 108 bpm and 118 bpm to avoid the situation where the columns occupied by an individual note would always be the same integer. 

\subsection{Dataset Formation}
Once converted to MIDI files, the computer-generated and human-composed melodies are divided into two datasets: the development dataset and the evaluation dataset. Neither datasets contain labels and they consist of an equal number of Bach-style and pop-style melodies.  

In the development dataset, there are 6,000 computer-generated melodies generated by three models. The specific composition of the development dataset is shown in Table~\ref{tab:developement}.

For each type of music generation system, two different datasets were used for training two individual melody generation systems: melodies from Bach Chorales in Music21 (labelled as ``Bach'' in Table \ref{tab:developement}) and hooktheory dataset (labelled as ``Pop'' in Table \ref{tab:developement}).

\begin{table}[htb]
	\centering
	\caption{The development dataset composition of the data challenge where the number in the brackets indicates the number of melodies. ``MTrans'', ``MVAE'' and ``MNet'' represent for music transformer, MusicVAE and MidiNet respectively. }
	\begin{tabular}{cccccc}
		\toprule[1pt]
		 \multicolumn{6}{c}{Computer-generated music (6000)} \\
		\midrule[1pt]
		\multicolumn{2}{c}{MTrans (2000)} & \multicolumn{2}{c}{MVAE (2000)} & \multicolumn{2}{c}{MNet (2000)} \\
		\midrule[0.5pt]
		Bach & Pop & Bach & Pop & Bach & Pop \\
        \midrule[0.5pt]
        1000 & 1000 & 1000 & 1000 & 1000 & 1000 \\
		\bottomrule[1pt]
	\end{tabular}
	\label{tab:developement}
\end{table}

In the evaluation dataset, there are 4,000 melodies coming from two sources: computer models and human composition. 

Among the human-composed melodies, the items truncated from melodies originally used for training music generation systems (labelled as ``Training'' in Table \ref{tab:evaluation}) and specially composed melodies for this data challenge (labelled as ``Unpublished'') are delineated given the two styles: from Bach Chorales or similar with Bach style (labelled as ``Bach'' in Table \ref{tab:evaluation}) and from hooktheory dataset or common pop style (labelled as ``Pop'' in Table \ref{tab:evaluation}).

The composition of computer-generated melodies are complex. As a general principle, it is necessary to emphasise that the system used to generate melodies in the evaluation dataset and the system used to generate melodies in the development dataset are always different although system architectures may be shared. As the case in the development dataset, each proposed system is trained using two different datasets (labelled as ``Bach'' and ``Pop'' in Table \ref{tab:evaluation}) hence two separate melody generation systems for different styles are obtained. 

Table \ref{tab:evaluation} summarises the composition of the evalution dataset. It is worth mentioning that numbers of melodies generated by MusicTransformer is larger than the other systems in order to investigate the effects of different initialisation configurations. Unlike in the development dataset where only one configuration used for initialisation of the MusicTransformer, the melodies in the evaluation dataset generated by MusicTransformer are the result of three different initialisation configurations, among which one of the initialisation scheme is used in the training process. 

\begin{table}[thb]
	\centering
	\caption{The evaluation dataset composition of the data challenge where the number in the brackets indicates the number of melodies. ``MTrans'', ``MVAE'' and ``MNet'' represent for music transformer, MusicVAE and MidiNet respectively. The number in the brackets indicates the number of melodies. The title of each column is explained in the context.}
	\begin{tabular}{cccccccccccc}
		\toprule[1pt]
		\multicolumn{12}{c}{Computer-generated melodies (2000)} \\
		\midrule[1pt]
		\multicolumn{4}{c}{MTrans (1200)} & \multicolumn{4}{c}{MVAE (400)} & \multicolumn{4}{c}{MNet (400)} \\ 
		\midrule[0.5pt]
		\multicolumn{2}{c}{Bach} & \multicolumn{2}{c}{Pop} & \multicolumn{2}{c}{Bach} & \multicolumn{2}{c}{Pop} & \multicolumn{2}{c}{Bach} & \multicolumn{2}{c}{Pop} \\
        \midrule[0.5pt]
        \multicolumn{2}{c}{600} & \multicolumn{2}{c}{600} & \multicolumn{2}{c}{200} & \multicolumn{2}{c}{200} & \multicolumn{2}{c}{200} & \multicolumn{2}{c}{200} \\
        \midrule[1pt]
        \multicolumn{12}{c}{Human-composed melodies (2000)} \\
        \midrule[1pt]
        \multicolumn{6}{c}{Training (1900)} & \multicolumn{6}{c}{Unpublished (100)} \\
        \midrule[0.5pt]
        \multicolumn{3}{c}{Bach} & \multicolumn{3}{c}{Pop} & \multicolumn{3}{c}{Bach} & \multicolumn{3}{c}{Pop} \\
        \midrule[0.5pt]
        \multicolumn{3}{c}{950} & \multicolumn{3}{c}{950} & \multicolumn{3}{c}{50} & \multicolumn{3}{c}{50} \\
		\bottomrule[1pt]
	\end{tabular}
	\label{tab:evaluation}
\end{table}

\section{Conclusions}
The CSMT data challenge requires participants to identify computer-generated melodies among human-composed melodies. The challenge aims to facilitate solutions for determining the source of melodies in possible copyright infringement cases in juridical practice. The term ``melody'' is used in a limited sense in this data challenge. Melodies were truncated to remove musical structure and they were used without accompaniment. This paper provided an in-depth discussion on the composition and the design of the dataset. 

The challenge utilises two components, the development dataset and the evaluation dataset. The development dataset contains only computer-generated melodies whereas the evaluation dataset combines both computer-generated and human-composed melodies. The computer-generated melodies in the development and evaluation datasets are obtained from the same type of systems with slightly different settings. The human-composed melodies were composed specifically for the CSMT data challenge besides existing melodies that were used for system training. 

With the presented setup of the challenge, the identification of computer-generated melodies can be considered either an unsupervised outlier detection problem or a supervised classification problem. Both methodologies may suffer from learning the inherent limitations of the selected music generation systems. As a result, the systems proposed by participants in the data challenge may not produce a universally valid approach to identify computer generated melodies, but rely on data distributions instead that characterise state-of-the-art music generation systems. Nevertheless, this approach can still demonstrate valuable for practical purposes, as in the legal context introduced earlier, if the models are kept up to date. Moreover, the melody complexity in this data challenge is reduced artificially hence the algorithms from participants may have limited generalisability. 

\section*{Acknowledgement}

The authors acknowledge the contribution from all members in the organisation committee (other than the authors): Prof. ZHANG Ru from Beijing University of Posts and Telecommunications, Dr. LI Zijin from China Conservatory of Music, Mr. ZHU Yidan from Beijing Acoustics Society, Mr. ZHOU Wei from Beijing Zhongwen (Shanghai) Law Firm.

%
%
%
\bibliographystyle{splncs04}
\bibliography{related_work_sub}

\end{document}